\documentclass[aps,prd,twocolumn,nofootinbib]{revtex4}

\usepackage{amsmath,amssymb,epsfig,color}

\newcommand{\be}{\begin{equation}}
\newcommand{\ee}{\end{equation}}
\newcommand{\ben}{\begin{eqnarray}\displaystyle}
\newcommand{\een}{\end{eqnarray}}

\begin{document}

\title{Giant Rings in the CMB Sky}

\author{Ely D. Kovetz}
\email{elykovetz@gmail.com}
\author{Assaf Ben-David}
\email{bd.assaf@gmail.com}
\author{Nissan Itzhaki}
\email{nitzhaki@post.tau.ac.il}
\affiliation{Raymond and Beverly Sackler Faculty of Exact Sciences,
School of Physics and Astronomy, Tel-Aviv University, Ramat-Aviv, 69978,
Israel}

\begin{abstract}

We find a unique direction in the CMB sky around which giant rings
have an anomalous mean temperature profile. This direction is in
very close alignment with the afore measured anomalously large bulk
flow direction.
Using Monte Carlo simulations, we estimate the significance of the
giant rings at the $3\sigma$ level and the alignment with the bulk flow at $2.5\sigma$.
We argue that a cosmic defect seeded by a
pre-inflationary particle could explain the giant rings, the large
bulk flow and their alignment.

\end{abstract}

\maketitle

\section{Introduction}

One of the key assumptions in modern cosmology is statistical
isotropy. The detailed data from the Wilkinson Microwave Anisotropy
Probe (WMAP) provides an opportunity to test this assumption. Indeed
many authors have studied this issue directly and indirectly using
various approaches and claimed the existence of a number of
anomalies in the data (see e.g.
\cite{Spergel:2003, Copi:2007, Hajian:2007pi, Bennett:1992, Hinshaw:1996, Tegmark:2003, Schwarz:2004, Land:2005, Vielva:2004, Mukherjee:2004, Cruz:2006apj, Cruz:2006mnras, Cruz:2007, Cruz:2008, Eriksen:2004, Hansen:2004, Eriksen:2007, Hoftuft:2009, Groeneboom:2009, Joshi:2009mj, Souradeep:2006dz}
and  \cite{Abramo:2010gk, Copi:2010na} for   recent reviews).

In this paper we propose another approach to test statistical
isotropy. We study how much giant rings in the Cosmic Microwave
Background (CMB) sky deviate from random behavior and estimate the
significance of the deviation. In section II we define the {\it
rings score} as a function of the direction the giant rings surround
and generate a rings score map for the masked Internal Linear
Combination (ILC) map. We show that the ILC rings score map has a
clear peak. We estimate the giant rings in the ILC map 
as a 3$\sigma$ deviation from $\Lambda$CDM.
Moreover, we find that the giant rings are aligned with another reported
$\Lambda$CDM anomaly \cite{Watkins, BulkFlow, Lavaux} in the form of a large
cosmic bulk flow.

In section III we discuss a cosmological scenario that could explain
the giant rings, the large bulk flow and their alignment.  It is
this cosmological scenario, which  involves the effects of a
pre-inflationary particle \cite{Sunny, PIP}, that actually motivated
us to look for these giant rings in the first place. Section IV is devoted to
discussion.

\section{Giant rings in the sky}

\begin{figure}
\begin{picture}(50,150)(50,0)
\vspace{0mm} \hspace{0mm} \mbox{\epsfxsize=50mm
\epsfbox{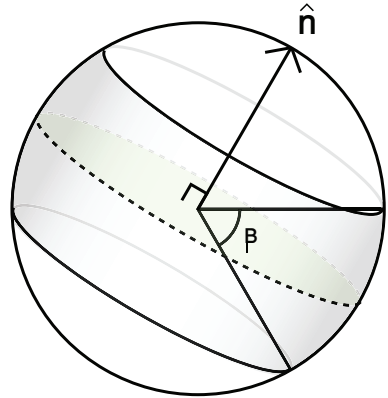}}
\end{picture}
\caption{The score is calculated for rings in a band of width $\beta$
surrounding the equatorial defined by a direction $\hat{n}$.\label
{fig:ScoreDrawing}}
\end{figure}

We begin  with the following question: Are there unusual rings in
the CMB sky? For reasons that will become clear shortly, we choose
to focus on the largest possible rings, namely those that reside in
a band of width $\beta$ around $\theta=\pi/2$ with respect to some
direction specified by a unit vector $\hat{n}$ (see
Fig.~\ref{fig:ScoreDrawing}).
The band is symmetric and so the range of angles considered is
\be \frac{\pi -\beta}{2}<\theta<\frac{\pi +\beta}{2}. \ee
We denote the average temperature of an infinitesimal ring by
$T(\theta, \hat{n})$  and the mean of the map by $T_0$ and  use the
following {\it rings score} to detect unusual rings in the sky
\be\label{con} R(\beta,\hat{n})=\int_{\frac{\pi
-\beta}{2}}^{\frac{\pi +\beta}{2}}d(\cos\theta)~ \tilde{
T}^2(\theta,\hat{n}),\ee
where $\tilde {T}(\theta,\hat{n})=T(\theta,\hat{n})-T_0$.  This
score is chosen since we are not looking to find any particular
shape of $\tilde{T}(\theta, \hat{n})$. Rather, we are searching for
the direction in which the rings deviate maximally from random
gaussian behavior. For this we simply need to weigh correctly the
contribution of each infinitesimal ring to our rings score. This is
the reason for the $d(\cos\theta)$ in the score.

There are some issues one needs to deal with when working with
actual CMB data. First, to have enough statistics in each
infinitesimal ring, the rings cannot be taken to be arbitrarily
small and the integral must be replaced by a discrete sum. In the
results reported below for the 7-year ILC map (given in $1^{\circ}$
resolution) we took $d\theta\rightarrow \Delta\theta= 3^{\circ}$,
but have verified that the results are not sensitive to
$\Delta\theta$. Secondly, for obvious reasons we would like the
results to be insensitive to Galactic foregrounds. Hence we use the
KQ75 mask which removes $29\%$ of the WMAP7 sky and  calculate the
quantity
\be\label{dis} R_{\text{dis}}(\beta,\hat{n})= \sum_{i=1}^{\beta/
\Delta\theta} \tilde{T}^2(i, \hat{n}) ~ M(i, \hat{n}) , \ee
where $M(i, \hat{n})$ is the number of pixels in the $i$'th ring
that survived the KQ75 mask cut and $\tilde{T}(i, \hat{n})$ is the
difference between the average temperature in the $i$'th ring around
the direction $\hat{n}$ and the mean of the masked map.

We would like to test the isotropy assumption of $\Lambda$CDM via
the rings score.  With this goal in mind  $\beta$ cannot  be taken
to be too small since this  will increase the chance that the
direction favored by $R_{\text{dis}}(\beta,\hat{n})$ has no
significant importance and is merely a statistical fluke. However,
due to the mask we are using we cannot take $\beta$ to be too large
either. The reason is that as we increase $\beta$ the average size
of an infinitesimal ring becomes smaller and so the ratio between
the number of pixels we are masking and the pixels we are keeping
when calculating $\tilde{T}(i, \hat{n})$ becomes larger for generic
values of $\hat{n}$.

Note that even if $\beta$ is small $R_{\text{dis}}(\beta,\hat{n})$
does not approximate $R(\beta,\hat{n})$ well when $\hat{n}$ points
roughly perpendicular to the galactic plane, because the
corresponding rings lie mainly inside the mask. To be on the safe
side, we ignore directions for which more than $80\%$ of the  rings
have less than $30\%$ unmasked pixels in each ring. Thus using the
cut sky eliminates a small portion of the map around the north and
south poles.

\begin{figure}[b!]
\includegraphics[width=\linewidth]{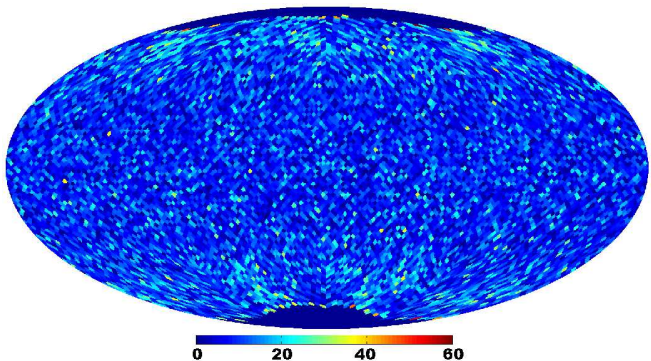}
\caption{A histogram showing the location of the maximum rings score (with $\beta=\pi/3$) for each of $100,\!000$ randomly generated maps using the KQ75 mask. The maximum scores for $27\%$ of the maps lie within the mask, which covers $29\%$ of the sky.\label{fig:HistRingScore}}
\end{figure}

To verify that the defined rings score is not biased by the mask we used 
the HEALPix package to generate $100,\!000$ random maps and masked them with the KQ75 map.
We found that the peaks in the corresponding rings score maps are
uniformly distributed on the sky and do not favor any particular area (see Fig.~\ref{fig:HistRingScore}).

\begin{figure}
\includegraphics[width=\linewidth]{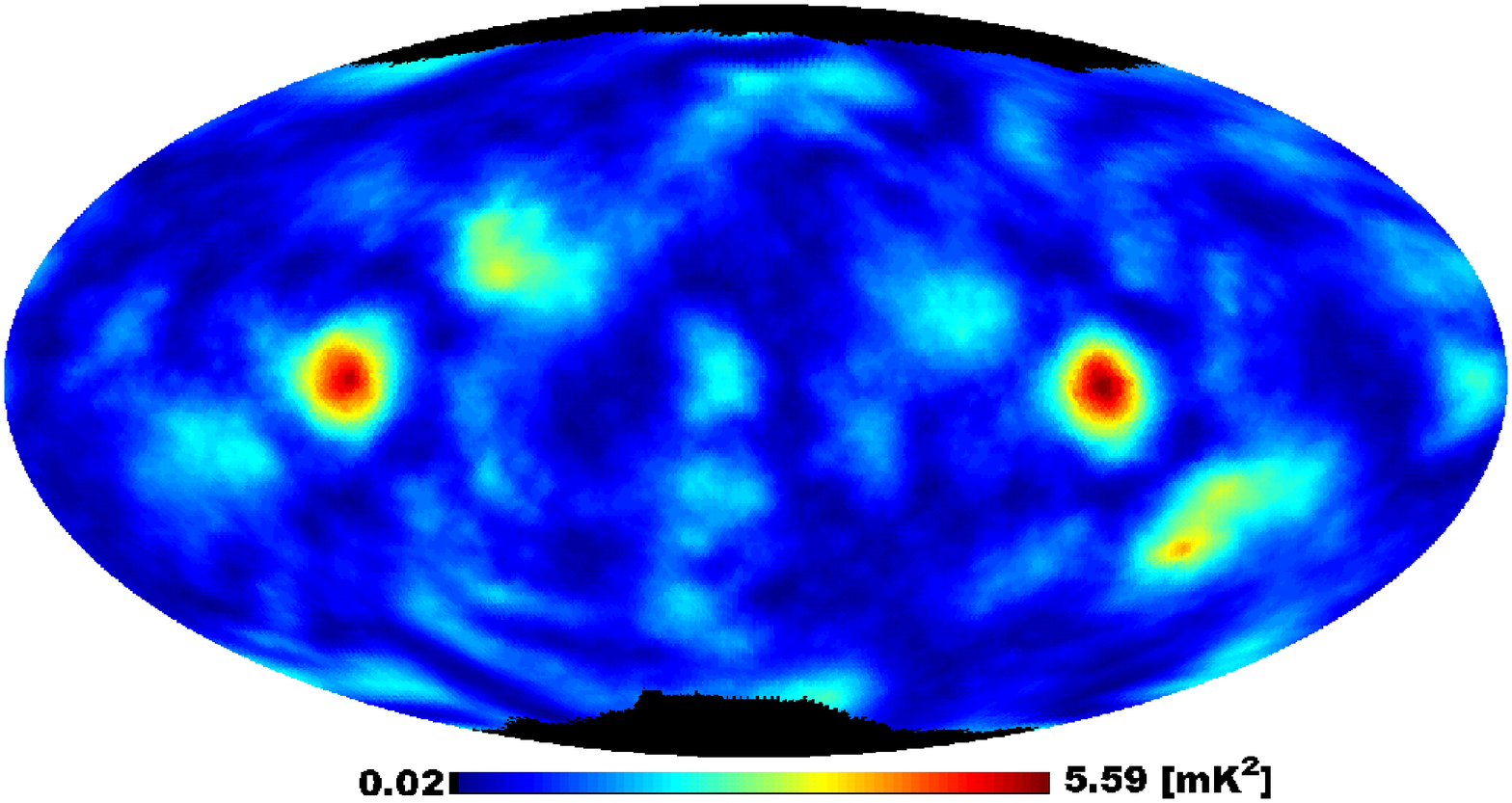}
\includegraphics[width=\linewidth]{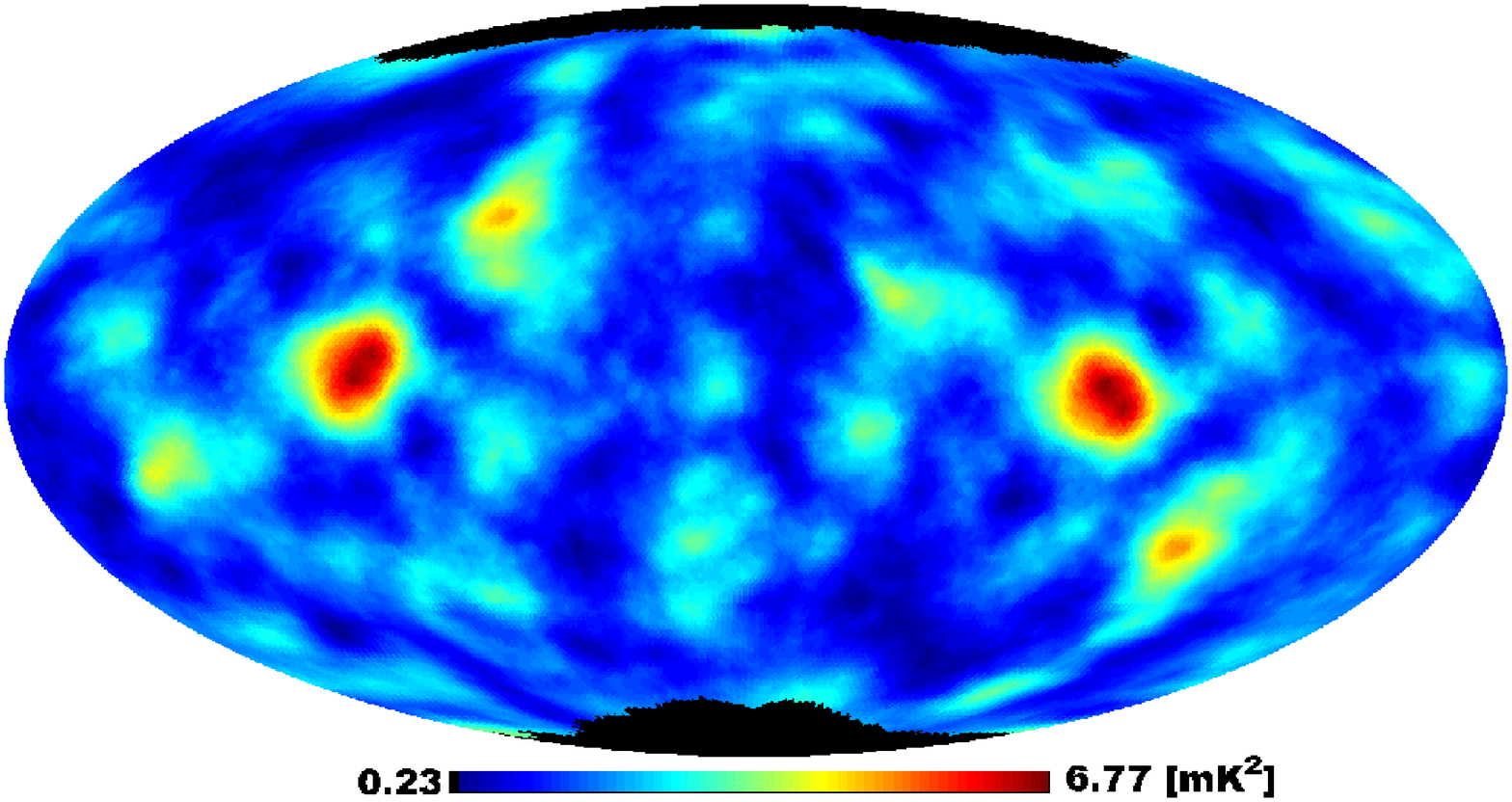}
\includegraphics[width=\linewidth]{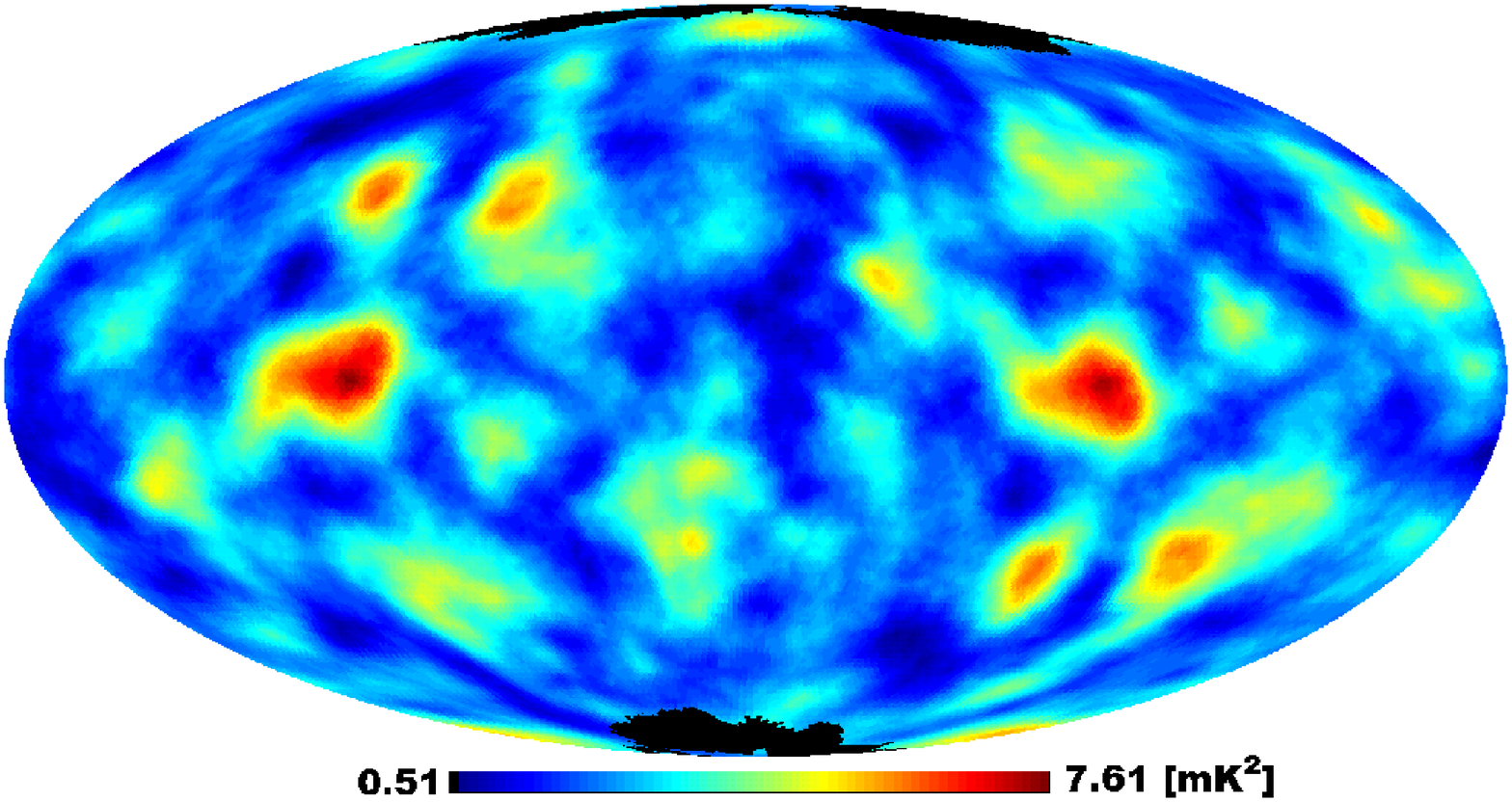}
\caption{\emph{From top to bottom:} The rings score calculated for $\beta=\pi/6, \pi/3$ and $\pi/2$, respectively. The score is calculated on the 7-yr WMAP ILC map (degraded to $N_{\text{side}}=64$) with the KQ75 mask applied. Ignored directions are marked in black.\label{fig:RingScore}}
\end{figure}

\begin{figure}
\includegraphics[width=\linewidth]{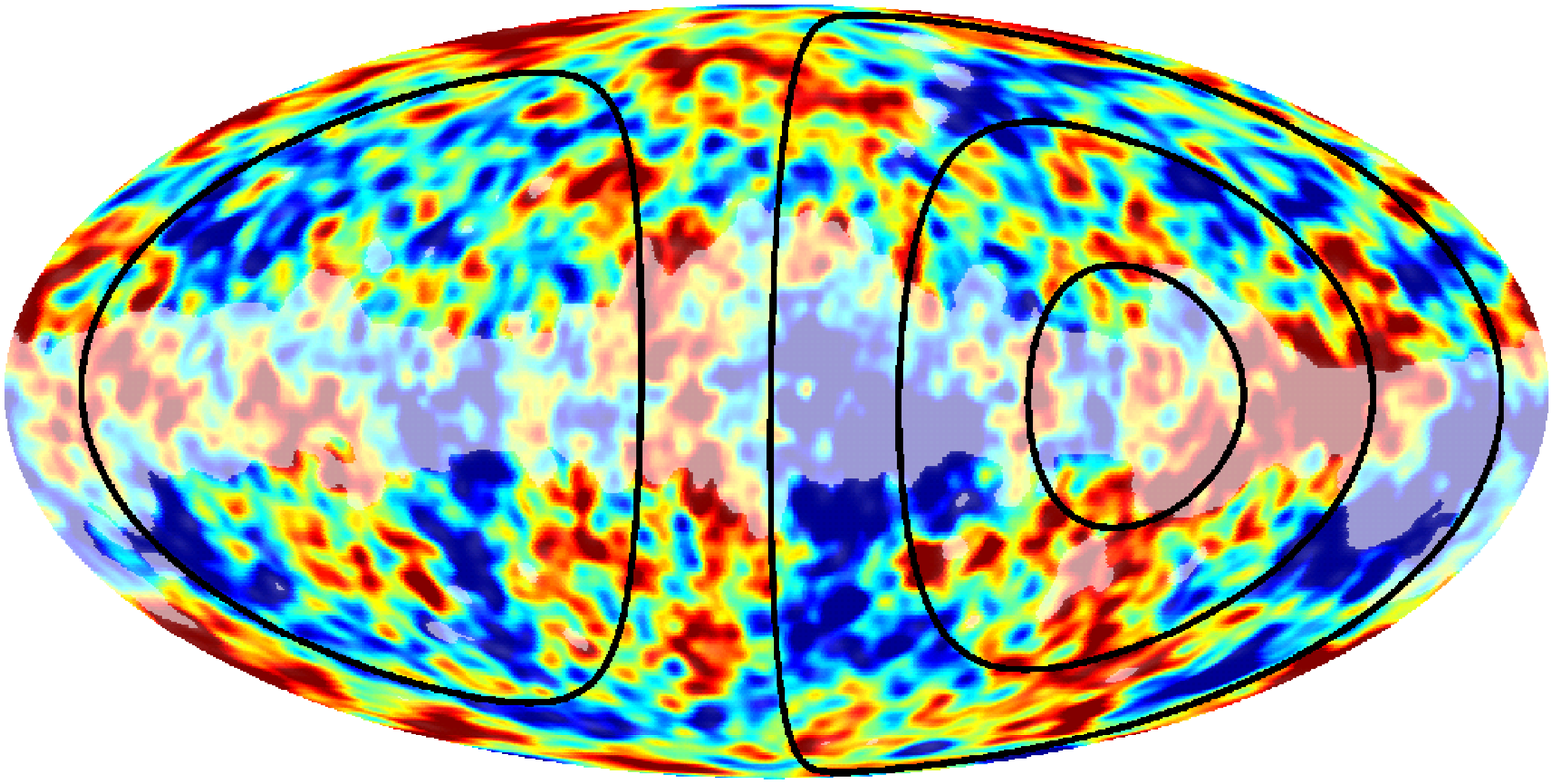}
\includegraphics[width=\linewidth]{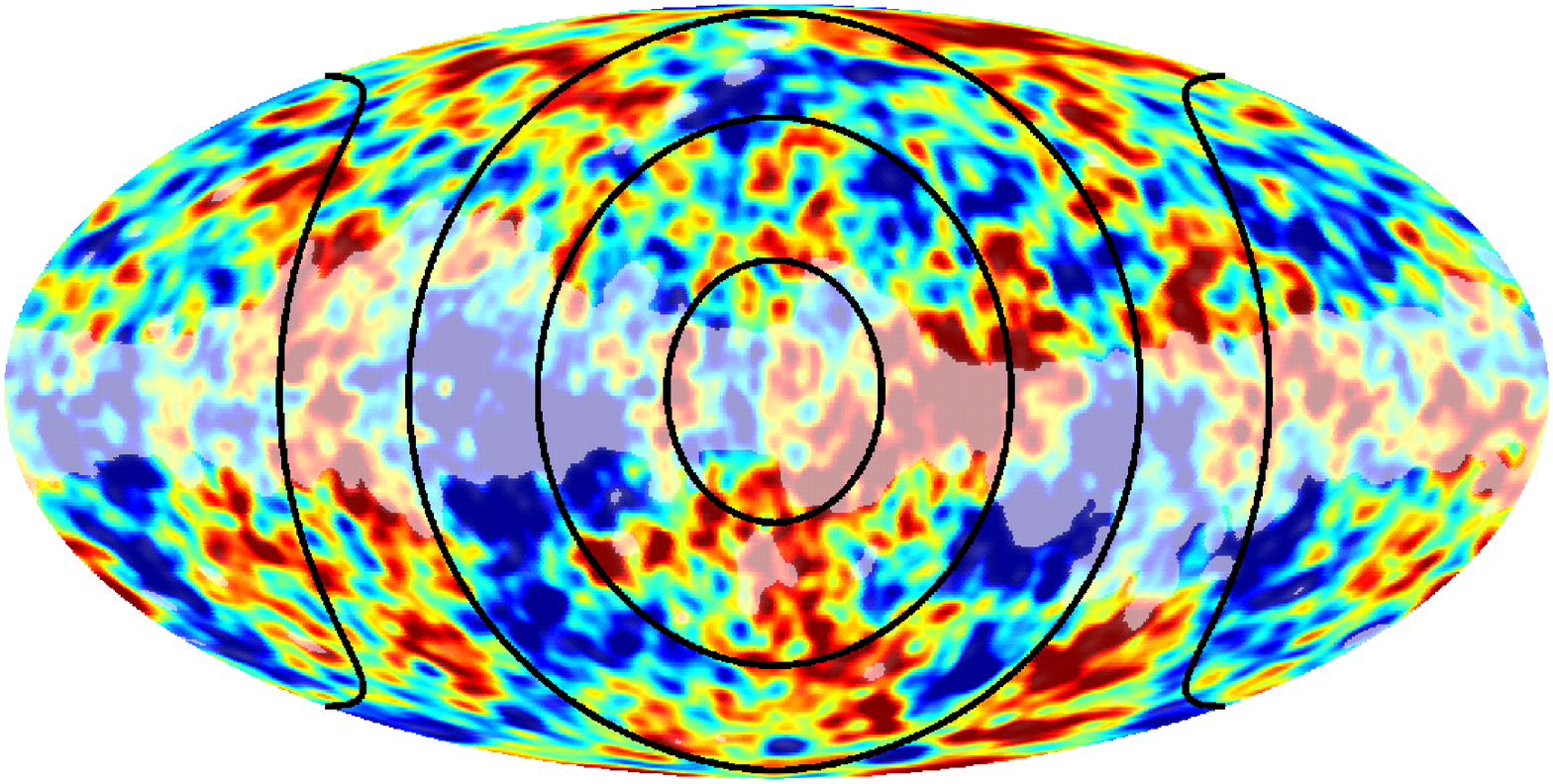}
\caption{\emph{Top:} The 7 year ILC map, smoothed to $3^{\circ}$ resolution.
The KQ75 mask is faintly superimposed on the map and the rings are
marked
around the dominant direction. \emph{Bottom:} The same map rotated so that
the dominant direction is placed at the center of the map. Incidentally, two of
the
\emph{Cold Fingers} discussed recently by the WMAP team \cite{WMAP7}
(the ones that include what \cite{WMAP7} refer to as {\it Cold Spot I} and {\it
Cold Spot
II}) fall nicely inside the cold ring ($55^\circ<\theta<85^\circ$), with their hot
counterparts falling inside two surrounding hot rings ($25^\circ<\theta<55^
\circ$ and
$85^\circ<\theta<115^\circ$).\label{fig:RingsVisibleInILC}}
\end{figure}

In Fig.~\ref{fig:RingScore} we plot the rings score
calculated for the 7-year ILC map with the KQ75 mask for $\beta=\pi/6, \pi/3$ and $\pi/2$. The dark
regions near the poles are ignored for the reason explained above.
We see that at around Galactic coordinates $(l,b) = (276^\circ,-1^\circ)$
there is a distinct peak.\footnote{Up to effects of the asymmetric masking the rings score is
symmetric under a $180^\circ$ inversion and so a symmetric partner peak
appears as well. Taking an asymmetric $\theta$ range, however, the direction
$(276^\circ,-1^\circ)$ is more prominent.}
In fact, as illustrated in Fig.~\ref{fig:RingsVisibleInILC},
even with the naked eye, the giant rings are visible.

There are some interesting aspects to this rings score map in general
and the peak at  $(276^\circ,-1^\circ)$ in particular. 
The location of the peak of $R_{\text{dis}}(\beta,\hat{n})$ is fairly insensitive to the value of $\beta$ (as long as it is not too small or too large). To be precise, when varying $\beta$ from $30^\circ$ to $165^\circ$, taking 10 evenly spaced values, the peak moves at most by $2^\circ$. Repeating this test on $10,\!000$ randomly generated maps (masked with the KQ75 mask), only in 14 cases the peak was as stable. This implies a 3$\sigma$ deviation from the statistically isotropic $\Lambda$CDM model.

As is evident from Fig.~\ref{fig:RingScore} most of the signal comes from the largest rings. 
To quantify this 
we calculate for each map ($\beta=\pi/6, \pi/3$ and $\pi/2$) the score
\be S(\beta) =\frac{R_{\max}-\bar{R}}{\sigma}, \ee
where $R_{\max}$ is the maximum of the rings score map and $\bar{R}$
and $\sigma$ are the mean and standard deviation of the rings score
map and get $S_\text{ILC}(\beta)=7.09, 5.85$ and $4.93$, respectively.
We estimate the significance by calculating $S(\beta)$ on random maps
and find that for $\beta=\pi/6, \pi/3$ and $\pi/2$ only $0.16\%, 0.57\%$ and $3.97\%$ of the maps get a higher score, respectively.
This implies that most of the signal comes from a narrow band around $\theta=\pi/2$, and that as we increase $\beta$ we increase the noise without increasing the signal.

\begin{figure}[h!]
\begin{picture}(50,80)(70,0)
\vspace{0mm} \hspace{0mm} \mbox{\epsfxsize=40mm
\epsfbox{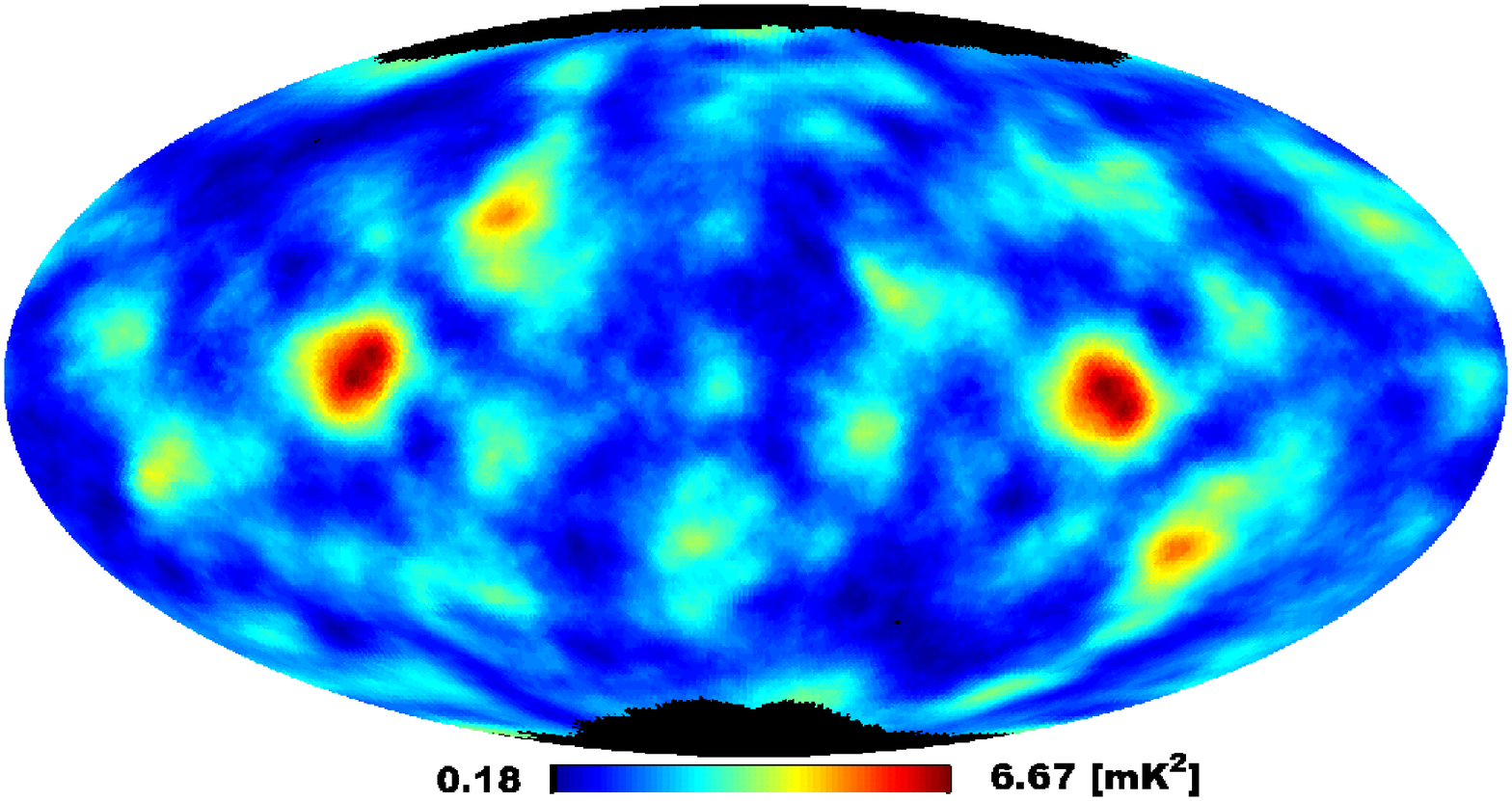}}
\end{picture}
\begin{picture}(50,80)(0,0)
\vspace{0mm} \hspace{0mm} \mbox{\epsfxsize=40mm
\epsfbox{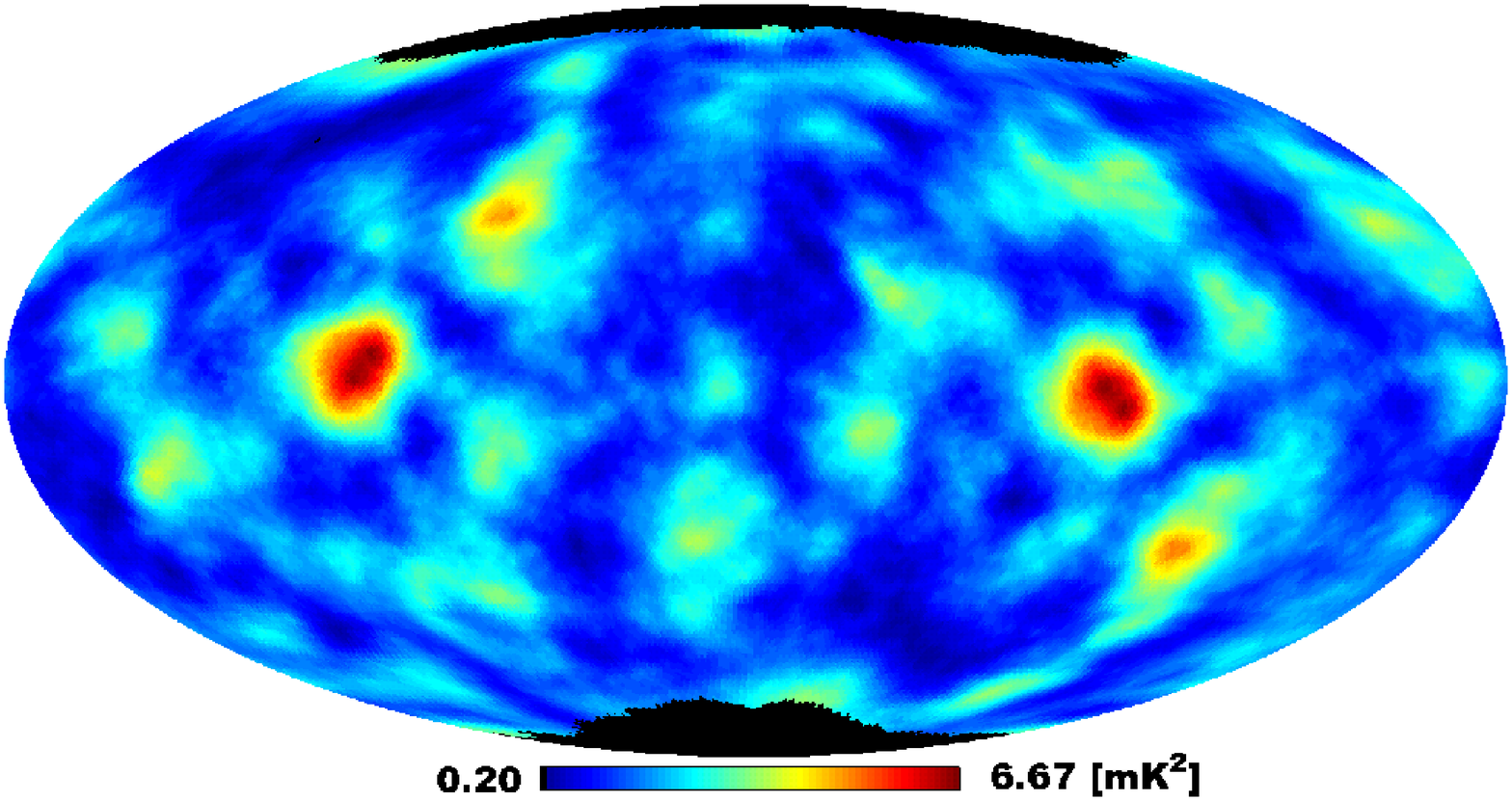}}
\end{picture}
\caption{$R_\text{dis}(\pi/3,\hat{n})$ for the
foreground reduced 7-yr WMAP V (\emph{left}) and W (\emph{right})
frequency band maps (degraded to $N_{\text{side}}=64$). The score maps
are very similar to one another and to the ILC score map.
In both maps, the peak is in the same location, $(276^\circ, -1^\circ)$ and its significance vs. random maps is $\sim1\%$. The location remains
the same also when tested in higher resolution of $N_\text{side}=128$.\label{fig:V and W}}
\end{figure}

Several features breaking statistical isotropy have been found in WMAP data that turned out to be the result of astrophysical or systematic effects and not of cosmological origin. Of course, we cannot completely rule out a similar explanation for the giant rings, but there is support for a cosmological explanation. First, the rings score maps for the V and W frequency bands are almost identical to that of the ILC (see Fig.~\ref{fig:V and W}). Secondly, the ecliptic pole is located at $(276^\circ,-30^\circ)$ and is $\sim30^\circ$ away from both the main rings score peak and the secondary peak at $(248^\circ,-34^\circ)$.

\begin{figure}
\includegraphics[width=\linewidth]{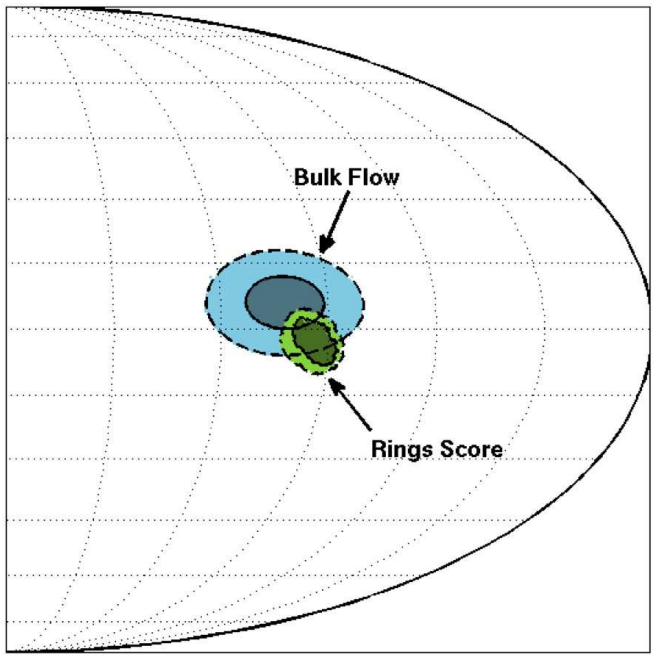}
\caption{The rings score direction $(276^\circ \pm 9^\circ,-1^\circ \pm 7^\circ)$ and the Bulk Flow direction $(282^\circ \pm 11^\circ, 6^\circ \pm 6^\circ)$. The $1\sigma$ distance between the two is $9^{+12}_{-9}$ degrees. Areas for $1\sigma$ and $2\sigma$ are shown.\label{fig:EllipseRingScore}}
\end{figure}

Further support for the cosmological origin of the giant rings is the intriguing alignment between the direction
of the rings and the direction of the
large bulk flow reported in \cite{BulkFlow}. According to \cite{BulkFlow}
the bulk flow on scales of about $100$ Mpc/h has a magnitude of $|v|
= 416 \pm 78$km/s towards $(l,b) = (282^\circ \pm 11^\circ, 6^\circ \pm 6^\circ)$.
The chance of such a large bulk flow to happen in $\Lambda$CDM on
such large scales is about $0.5\%$ \cite{BulkFlow}.
As plotted in Fig.~\ref{fig:EllipseRingScore}, for $\beta=\pi/3$ the rings score direction is $(276^\circ \pm 9^\circ,-1^\circ \pm 7^\circ)$ and the distance between it and the Bulk Flow direction is $9^{+12}_{-9}$ degrees. The probability of a $9^\circ$ alignment between two random axes in the sky is $1.3\%$.

Overall, with current data we have two large scale quantities (one
is large scale in terms of the CMB and the other in terms of Large
Scale Structure) that are slightly anomalous 
that point roughly to the same direction.  In $\Lambda$CDM
there is no correlation between the bulk flow and the rings score
and so a fair point of view is to attribute the alignment between
the two to a statistical fluke (which is not so rare -- a $1.3\%$
effect) and to argue that both features are not anomalous enough  to
challenge $\Lambda$CDM.

In the next section we would like to offer a different point of
view. We show that the scenario of \cite{Sunny, PIP} naturally
explains the anomalous bulk flow,  the giant rings and their
alignment. Taking the above results at face value, this scenario
explains a one-in-a-million effect in $\Lambda$CDM.

\section{A possible cosmological explanation}

The fact that the bulk flow appears to be too large is known for
quite some time now. What is fairly new is that the shear and
octupole moments associated with the bulk motion appear to be
consistent with $\Lambda$CDM \cite{BulkFlow} (see also \cite{Song:2010kq}). 
This suggests that these higher moments are generated by the standard $\Lambda$CDM
power spectrum, while the overall bulk flow is generated by a
non-$\Lambda$CDM  faraway source.

A model with exactly this feature (motivated by \cite{StringTheory}
and earlier works \cite{e1, e2, e3}) was suggested recently in
\cite{Sunny, PIP}. There, we studied  some of the cosmological
imprints of pre-inflationary particles (PIP). We found that each PIP
provides the seed for a giant structure (a spherically symmetric
Cosmic defect -- SSCD) whose gravitational potential is determined
by the PIP in the following way
\be \Phi_0(k)=\left.\frac{\lambda H}{12 \sqrt{\pi \epsilon}
k^3}\right|_{k=a(t) H},\ee
where $\epsilon$ is the slow roll parameter, $\lambda= dm/d\phi
-m\sqrt{\epsilon/2} $ (with $\phi$ being the inflaton field) and, as
usual, the effect is evaluated at horizon crossing.

For simplicity and concreteness it was assumed in \cite{PIP} that $d
\lambda/ d\phi$ and $n_s-1$ are negligible to find
\be\label{log} \Phi(r,z=0)=\lambda C \log(r), ~~~~C=1.09 \times
10^{-5}. \ee
Relaxing these assumptions one typically finds
\be\label{alpha} \Phi(r) \sim r^{\alpha},\ee
with $|\alpha |\ll 1$.

Both (\ref{log}) and (\ref{alpha}) vary very slowly over large distances and therefore 
are quite different than typical
potentials generated by the $\Lambda$CDM power spectrum. As a result, such a SSCD has distinct
cosmological imprints \cite{PIP}. In particular, it can induce a large bulk flow
from far away
towards its direction while having negligible
effect on higher moments of the bulk motion. Hence it fits neatly with the
observations of \cite{BulkFlow}.

The CMB signal of a single SSCD (seeded by a PIP) is affected by its
magnitude ($\lambda$ in the case of (\ref{log})) and its distance
from the observer, denoted by  $z_0$. Setting the magnitude to
produce the measured bulk flow for each $z_0$, we remain with $z_0$
as a single free parameter. The CMB signal is made up of competing
contributions from the Sachs-Wolfe (SW) and the late integrated SW (ISW)
effects and as was shown in \cite{PIP}, it should be detectable in
the CMB in the sense that it is larger than the noise. However, near
$z_0\sim 3$  the two nearly cancel out, so that a SSCD located there
can account for the measured bulk flow while adding a low, but
detectable, signal to the CMB sky that would not immediately stand
out as an obvious violation of statistical isotropy.

So how does one search for the SSCD in the CMB data? In particular,
we wish to find a way to tell apart the CMB signal of a SSCD from
that of an unusually strong structure generated by the  $\Lambda$CDM
 power spectrum.

An overdense $\Lambda$CDM structure will induce a cold {\it spot} in the CMB sky if located at the last scattering surface and a hot {\it
spot} if located nearby. Because of the unique large distance
behavior of (\ref{log}) (or (\ref{alpha})), a SSCD will induce a more
complex imprint that spreads all over the CMB sky (see \cite{PIP}). The shape of this imprint is azimuthally symmetric and its profile depends on $z_0$. Therefore its generic signature is a spot surrounded by rings. An interesting case happens where the contributions of the SW and ISW effects almost cancel out in the low multipoles (the cancellation happens at a different $z_0$ for each multipole). This can lead to the disappearance of the spot for certain $z_0$ values.
Since the potential falls slowly with distance, the fact that the
circumference is maximal at $\theta=\pi/2$ dominates and so a
generic signature is the appearance of anomalous rings around
$\theta\sim\pi/2$ from the location of the PIP. Focusing on these
rings lowers the possibility that the score will confuse an atypical
$\Lambda$CDM structure with a SSCD seeded by a PIP.

Moreover, if indeed a single SSCD is responsible for most of the
bulk velocity observed in \cite{BulkFlow}, then it should be located
very near to the Galactic plane, where the small $\theta$ signal
will be contaminated by Galactic foregrounds much more than the
large $\theta$ profile. Even though part of the signal predicted by
the SSCD lies in small angles, limiting the search and focusing only
on large angles yields a signal that is cleaner both with respect to
Galactic foreground and ordinary $\Lambda$CDM effects.

For these reasons we defined the azimuthally invariant rings score
the way we did in the previous section. It is designed  to detect a
SSCD seeded by a PIP.

\begin{figure}[b!]
\includegraphics[width=\linewidth]{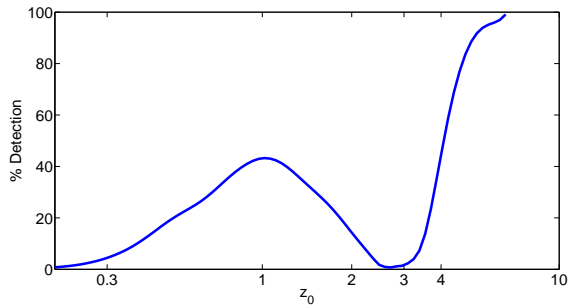}
\caption{Percentage of detection (to within a distance of $8^\circ$) of a
hidden SSCD at different redshifts $z_0$ by the rings score.\label
{fig:RingScoreSignal}}
\end{figure}

To verify that the score works as it should, we simulated random
CMB maps that include the contribution from a SSCD located at some
specific direction at a certain distance, and checked if the
direction with the maximal rings score is indeed near the SSCD
direction.
This is done by calculating the SSCD temperature imprint from the SW and ISW effects (taken from  \cite{PIP}) and adding it to the randomly generated map (we calculate the effect in a multipole expansion and add the first non-vanishing 10 multipoles $a_{\ell 0}$s to those of the random maps). The map is then rotated so that the SSCD is {\it hidden} at a desired direction. Next, we calculate the rings score on the map and check whether this direction indeed dominates the map.
In Fig.~\ref{fig:RingScoreSignal} we plot the percentage
of random maps with a hidden SSCD where the maximal rings score is less than $8^{\circ}$ away from the hiding location. We see that, as expected, it resembles the S/N graph in \cite{PIP} (Fig.~6(b)). A detailed study of the relation between the giant rings, other CMB anomalies and the PIP model will appear in a future work.

\section{Conclusions and Discussion}

In this paper we reported on a novel unexpected feature of the CMB
sky -- giant anomalous rings. The significance of these giant rings
by themselves, much like other ``anomalous" features of the CMB, is
far from being overwhelming as it is merely a $3\sigma$ effect.
Moreover, much like the absence of large angular correlations
\cite{Spergel:2003, Copi:2007}, our findings are weaker when considering the full ILC map with no mask -- the peak in the rings score map remains aligned with the bulk flow direction, but the score is no longer significant vs. random maps. This could either mean that this is due to the contamination of the unmasked data or that the feature is weaker than suggested by the masked map. With data
from Planck we should be able to determine which possibility is right. In
fact, since Planck is expected to clean much of the Galactic
foreground, we should be able to include the small $\theta$ data as
well and see if the signal increases.

Estimating the significance of features in the data is tricky in
general and in the case of statistical isotropy in particular.
Indeed many of the reported large scale anomalies in the CMB that
imply violation of statistical isotropy were recently deemed as
stemming from a-posteriori choices of estimators \cite{WMAP7,
Efstathiou:2009, Groeneboom:2009apj, Hanson:2009, Zhang:2009} that
amplified the significance of the results. Among the claims against
these anomalies is that they surfaced from a search of oddities in
the data with no independent experimental evidence or prior
theoretical motivation.

The giant rings are different  in this regard. First, the search for
them was motivated by a theoretical scenario which by construction
violates statistical isotropy. Secondly, they are aligned with
another large scale {\it non } CMB ``anomaly"  -- the bulk flow. This
increases, in our opinion, the chance that our findings could
eventually lead to a real challenge for statistical isotropy.
For this to happen more data is needed.

Luckily there are two clear predictions of our scenario that could
be tested already with Planck data. First, the weak gravitational
lensing of the CMB by the SSCD (assuming it has the magnitude
required to produce the large bulk flow) is quite distinct \cite{ab}
(again because of the long range gravitational potential it induces)
and should be detected by Planck. Secondly, the measurement of
peculiar velocities via the  kinetic Sunyaev-Zel'dovich effect
should improve quite significantly with Planck. This should enable
testing the claims of \cite{Kashlinsky}, which are based on data
from WMAP, and determine if the bulk flow is indeed anomalous at
even larger scales (which will increase the significance
dramatically), and in which direction it points. If our scenario is
correct then as one increases the size of the survey, the usual
$\Lambda$CDM effects should become more negligible compared to the SSCD effect and the measured bulk flow will be more aligned with the giant rings.

\acknowledgments We thank A. Fialkov and B. Keren-Zur for useful
discussions. We acknowledge the use of the Legacy Archive for
Microwave Background Data Analysis (LAMBDA) \cite{lambda} and the
use of the HEALPix package \cite{Gorski:2004by}. This work is
supported in part by the Israel Science Foundation (grant number
1362/08) and by the European Research Council (grant number 203247).

\end{document}